\documentclass[aps,prl,twocolumn,a4paper,superscriptaddress]{revtex4-1}
\usepackage{amssymb}
\usepackage[utf8]{inputenc}
\usepackage{graphicx}
\usepackage{bm,color,subfigure,amsmath}

\begin{document}
\title{Confined dissipative droplet solitons in spin-valve nanowires with perpendicular magnetic anisotropy}

\author{Ezio Iacocca}
\email{ezio.iacocca@physics.gu.se}
\affiliation{Physics Department, University of Gothenburg, 412 96, Gothenburg, Sweden}

\author{Randy K. Dumas}
\affiliation{Physics Department, University of Gothenburg, 412 96, Gothenburg, Sweden}

\author{Lake Bookman}
\affiliation{Department of Mathematics, North Carolina State University, Raleigh, North Carolina 27695, USA}

\author{Majid Mohseni}
\affiliation{Material Physics, School of ICT, Royal Institute of Technology, Electrum 229, 164 40, Kista, Sweden}

\author{Sunjae Chung}
\affiliation{Physics Department, University of Gothenburg, 412 96, Gothenburg, Sweden}
\affiliation{Material Physics, School of ICT, Royal Institute of Technology, Electrum 229, 164 40, Kista, Sweden}

\author{Mark A. Hoefer}
\affiliation{Department of Mathematics, North Carolina State University, Raleigh, North Carolina 27695, USA}

\author{Johan \AA{}kerman}
\affiliation{Physics Department, University of Gothenburg, 412 96, Gothenburg, Sweden}
\affiliation{Material Physics, School of ICT, Royal Institute of Technology, Electrum 229, 164 40, Kista, Sweden}

\begin{abstract}
Magnetic dissipative droplets are localized, strongly nonlinear dynamical modes excited in nano-contact spin valves with perpendicular magnetic anisotropy. These modes find potential application in nanoscale structures for magnetic storage and computation, but dissipative droplet studies have so far been limited to extended thin films. Here, numerical and asymptotic analyses are used to demonstrate the existence and properties of novel solitons in confined structures.  As a nanowire's width is decreased with a nano-contact of fixed size at its center, the observed modes undergo transitions from a fully localized two-dimensional droplet into a two-dimensional droplet edge mode and then a pulsating one-dimensional droplet.  These solitons are interpreted as dissipative versions of classical, conservative solitons, allowing for an analytical description of the modes and the mechanisms of bifurcation.  The presented results open up new possibilities for the study of low-dimensional solitons and droplet applications in nanostructures.
\end{abstract}
\maketitle

Spin transfer torque (STT) induced excitations in magnetic systems~\cite{Slonczewski1996,Berger1996,Ralph2008} have attracted significant attention during the past decade due to their interesting fundamental properties and potential for technological impact. STT can, e.g., be exerted by a spin-polarized current impinging upon a magnetic thin film or by current flow within a magnetically inhomogeneous sample~\cite{Berger1984}. The former is achieved in devices known as spin valves (SV)~\cite{Tsoi1998,Kiselev2003,Silva2008,Houssameddine2008}, where two magnetic layers are separated by a non-magnetic spacer. One of the magnetic layers is considered largely insensitive to external excitations (fixed), and spin-polarizes the applied dc current. The second (free) magnetic layer is subjected to STT, giving rise to current-tunable dynamical modes such as propagating spin-waves~\cite{Slonczewski1999,Pufall2006,Madami2011,Bonetti2010,Dumas2013}, localized spin-wave bullets~\cite{Slavin2005,Bonetti2010,Bonetti2012,Demidov2010,Demidov2012,Dumas2013}, vortices~\cite{Pufall2007,Mistral2008,Petit2012}, and dissipative droplets.~\cite{Hoefer2010,Hoefer2012,Mohseni2013} The required high current densities are usually achieved by patterning a nano-contact (NC) on top of the spin valve (NC-SV). On the other hand, STT induced by currents within a magnetically inhomogeneous sample provides the basis for current-induced domain-wall motion (CIDWM)~\cite{Boulle2011,Miron2011}, as demonstrated in nanowire ferromagnetic thin films~\cite{Yamaguchi2004,Vernier2004,Boone2010} and SVs~\cite{Ono1998,Khvalkovskiy2009}.

For both types of STT excitations, perpendicular magnetic anisotropy (PMA) materials are of fundamental and technological interest. PMA materials support topological (Skyrmions)~\cite{Komineas2005,Moutafis2009} and non-topological (magnetic dissipative droplets)~\cite{Hoefer2010,Mohseni2013} solitons which offer a novel approach to applications~\cite{Allwood2005,Xu2008,Parkin2008,Hayashi2008} and control methods~\cite{Hoefer2012,Hoefer2012b,Emori2013,Fert2013}. Particularly, the magnetic dissipative droplet (`droplet' in the following) features a dynamical nature recently studied theoretically~\cite{Hoefer2010,Hoefer2012,Hoefer2012b} and experimentally~\cite{Mohseni2013}. However, these studies have considered only two-dimensional (2D) thin films, so it is natural to inquire about the role of physical confinement that nanoscale applications would introduce. For this, we investigate the effect of lateral confinement on droplets with micromagnetic simulations and asymptotic methods.

This Letter describes how a droplet undergoes transitions to an edge droplet and then to a quasi-one-dimensional (quasi-1D) droplet as an extended magnetic thin film is reduced to a nanowire. The edge droplet is essentially a non-topological half 2D droplet, exhibiting a larger footprint with respect to the NC. On the other hand, the quasi-1D droplet establishes oscillating magnetic boundaries along the physically extended dimension of the nanowire and it is found to acquire a chirality consistent with soliton-soliton pairs that exhibit breathing, similar to 1D solitons in biaxial ferromagnets~\cite{Kosevich1990,Braun2012}. Owing to the abovementioned features, the mode transitions are evidenced by distinct precessional frequencies. The presented results suggest these novel droplets as candidates for low-dimensional applications and soliton research in an experimentally realizable system.

\begin{figure}[t]
\centering\includegraphics[width=2.3in]{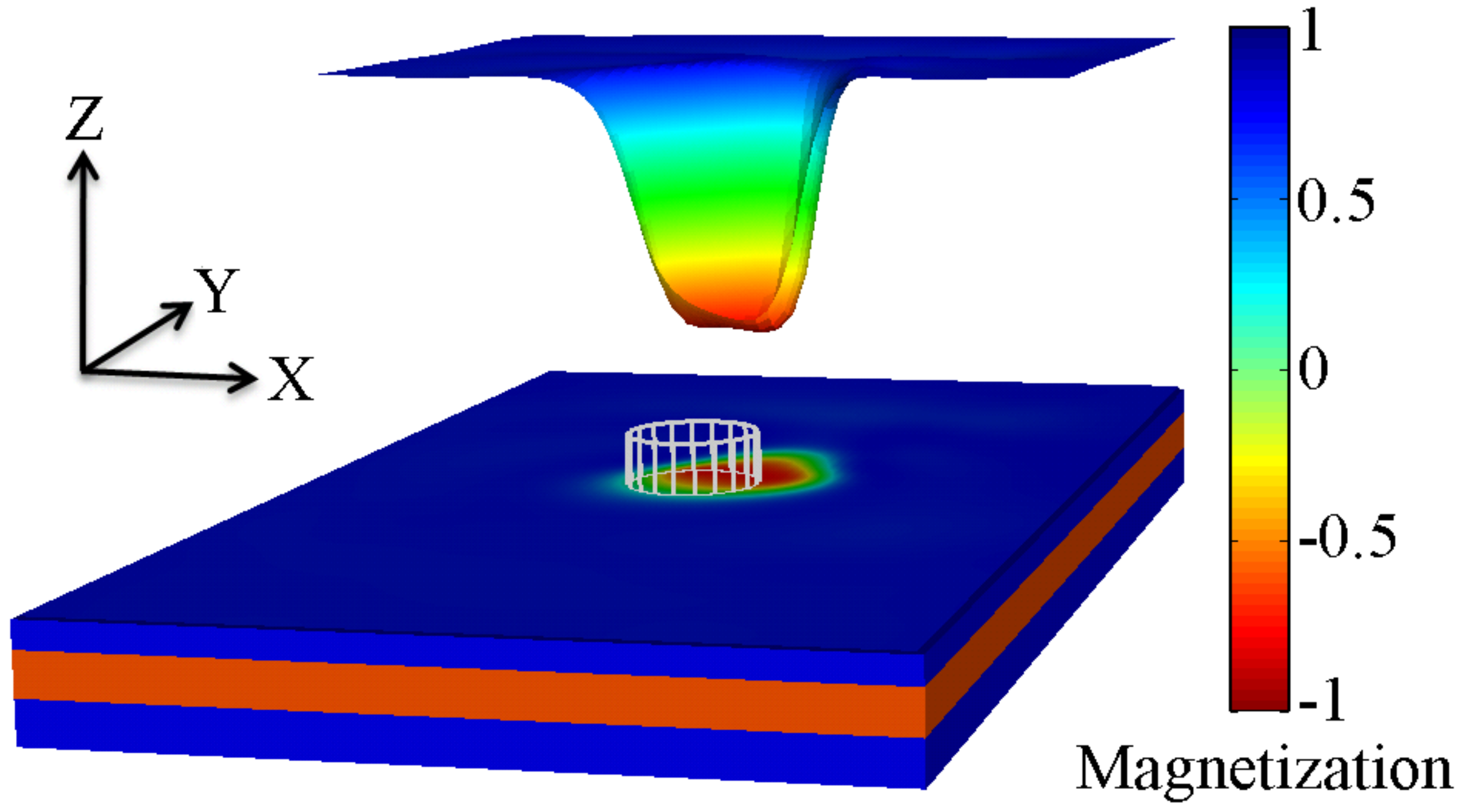}
\caption{ \label{fig1} 2D droplet excited in an extended spin-valve, showing its mostly reversed core. Below, the schematic of the spin-valve is shown, where the non-magnetic spacer is identified in ochre while the PMA material coloring follows the $m_z$ component as indicated in the colorbar. The nano-contact is placed in the geometrical center of the simulated area. }
\end{figure}

Droplets are two-dimensional, non-topological modes sustained by the STT-induced creation of an effective zero damping region below the NC, i.e., a gain - loss balance. In such a region, the magnetization is mostly reversed, creating a \emph{dynamic} magnetic boundary with its environment (Fig.~\ref{fig1}). This boundary defines the droplet size and it is expected to be strongly affected by lateral confinement. In order to explore such consequences, we begin by performing micromagnetic simulations of nanowires of decreasing width while keeping the NC laterally centered and of fixed radius. The droplet nucleation is due to a spin-wave modulational instability~\cite{Hoefer2010} leading to strongly nonlinear dynamics, often requiring micromagnetic simulations to uncover their features. However, an analytical treatment is available when some simplifications are made, as will be discussed below.

The system under study is a trilayered NC-SV consisting of PMA free and fixed layers (Fig.~\ref{fig1}). Micromagnetic simulations are performed for the free layer with the GPU-based tool Mumax2~\cite{Vansteenkiste2011}, using a second-order Runge-Kutta solver with an adaptive step bounded between $1$~fs and $1$~ps~\cite{Dumas2013}. The dynamics follow the Landau-Lifshitz-Gilbert-Slonczewski (LLGS) equation
\begin{equation}
\label{eq:llgs}
   \frac{d\hat{m}}{dt} = -\gamma\hat{m}\times\vec{H}_{eff}+\alpha\hat{m}\times\frac{d\hat{m}}{dt}-\gamma \mu_0 \sigma(I) f(\vec{x})\epsilon \hat{m}\times\hat{m}\times\hat{M},
\end{equation}
where $\gamma=28$~GHz/T is the gyromagnetic ratio, $\hat{m}$ and $\hat{M}$ are the normalized free and fixed layer magnetization vectors, respectively, $\alpha$ is the Gilbert damping, and $\sigma(I) = \hbar I P \lambda / \mu_0 M_S^2 e V(\lambda+1)$ is the dimensionless spin torque coefficient where $\hbar$ is the reduced Planck constant, $I$ is the spin-polarized current, $P$ is the polarization, $\epsilon=1/(1+\nu\hat{m}\cdot\hat{M})$, $\nu = (\lambda-1)/(\lambda+1)$, $\lambda$ is the spin torque asymmetry, $\mu_0$ is the vacuum permeability, $M_S$ is the saturation magnetization, $e$ is the electron charge, and $V$ is the free layer volume. STT is active only within the NC, a disk centered at the origin with diameter $d$ defined by $f(\vec{x}) = 1$, $|\vec{x}| < d/2$ and zero elsewhere. The effective field $H_{eff}$ includes the exchange, demagnetizing, anisotropy, and an out-of-plane external field $\mu_0H_a = 0.4$~T. The current-induced Oersted field is also included in the simulations by approximating the current flow path as an infinite cylinder. We assume material parameters measured on similar Co/Ni multilayers~\cite{Mohseni2011,Mohseni2013}, namely: thickness $5$~nm, $\alpha=0.05$, magnetic anisotropy $K_u=447$~kJ/$\mathrm{m^3}$, $M_S=716.2$~kA/m, and exchange stiffness $A=30$~pJ/m -- similar to Co --. The current-polarizing fixed layer is assumed to be perfectly out-of-plane with spin torque asymmetry $\lambda=1.1$ and polarization $P=1$ for simplicity. The unitary polarization generally leads to an underestimation of the current-induced Oersted field. However, its effect on the presented results is negligible~\cite{SuppMat2013} and it is only included for completeness. All simulations are performed in the absence of thermal fluctuating fields unless specified.

In order to study the effect of lateral confinement, the simulated layer's length (\textbf{x} axis) is fixed to $1000$~nm while the width (\textbf{y} axis) is varied between $300$ and $50$ nm. The cell-size is determined for each nanowire width to boost performance, yet defining an upper limit of $5$~nm, below the exchange length $\lambda_{ex}\approx 8.2$~nm. The NC has a diameter $d = 50$~nm (Fig.~\ref{fig1}) from which current is assumed positive when flowing from the free to the fixed layer. Note that the different cell-sizes introduce a discretization error in the STT region, $f(\vec{x})$. However, the robustness of the presented dynamics indicates that such error is negligible.

\begin{figure}[t]
\centering \includegraphics[width=3.3in]{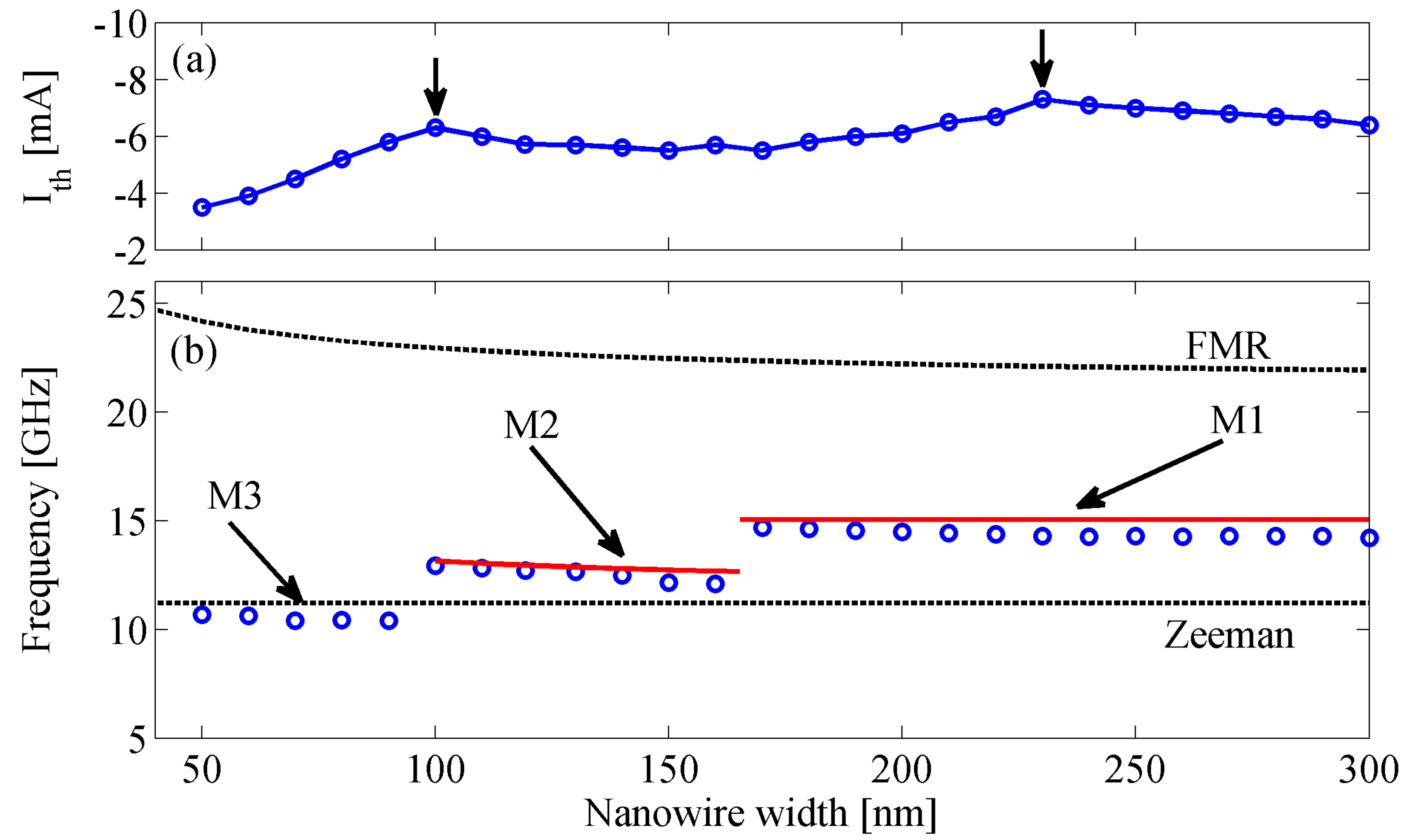}
\caption{ \label{fig2} (a) Threshold current for droplet nucleation and (b) droplet frequency at threshold as a function of the nanowire width. Three modes are identified: dissipative (M1), edge (M2), and quasi-1D (M3) droplet. The FMR and Zeeman frequencies are shown in black dashed lines. The analytically calculated frequencies are shown in solid red lines. }
\end{figure}

Droplets of a varying nature are excited above the threshold currents shown in Fig.~\ref{fig2}(a), determined by the condition $\langle m_z(t>10~\mathrm{ns}) \rangle_{NC}<0$, where $\langle \cdot \rangle_{NC}$ indicates average under the NC, while sweeping the current in steps of $0.1$~mA. This criterion is motivated by the fact that 2D droplets always contain a significant portion of the magnetization pointing to the southern hemisphere~\cite{Hoefer2010}. Fig.~\ref{fig2}(a) exhibits current local maxima, indicated by arrows, related to standing spin-waves across the nanowire width which preclude the onset of the modulation instability that gives rise to droplets~\cite{Hoefer2010}. Consequently, higher currents are needed to induce the spin-wave amplitude growth~\cite{SuppMat2013}.

The droplet frequencies are determined from $200$~ns time-traces sampled at $\sim 10$~ps [Fig.~\ref{fig2}(b)], from which three different modes are identified.  The frequencies for M1 and M2 lie between the $\mathrm{FMR}=\gamma\mu_0(H_a+H_k-N_zM_S)$ and $\mathrm{Zeeman}=\gamma \mu_0H_a$ frequencies -- both shown as black dashed lines -- where $N_z$ is the out-of-plane demagnetizing factor and $\mu_0H_k=2K_u/M_S=1.25$~T is the anisotropy field~\cite{Mohseni2011,Mohseni2013}. These frequency bounds coincide with those for 2D droplets in extended thin films~\cite{Hoefer2010}. Contrarily, M3 exhibits a sub-Zeeman frequency.

\begin{figure}[t]
\centering
\includegraphics[width=3.3in]{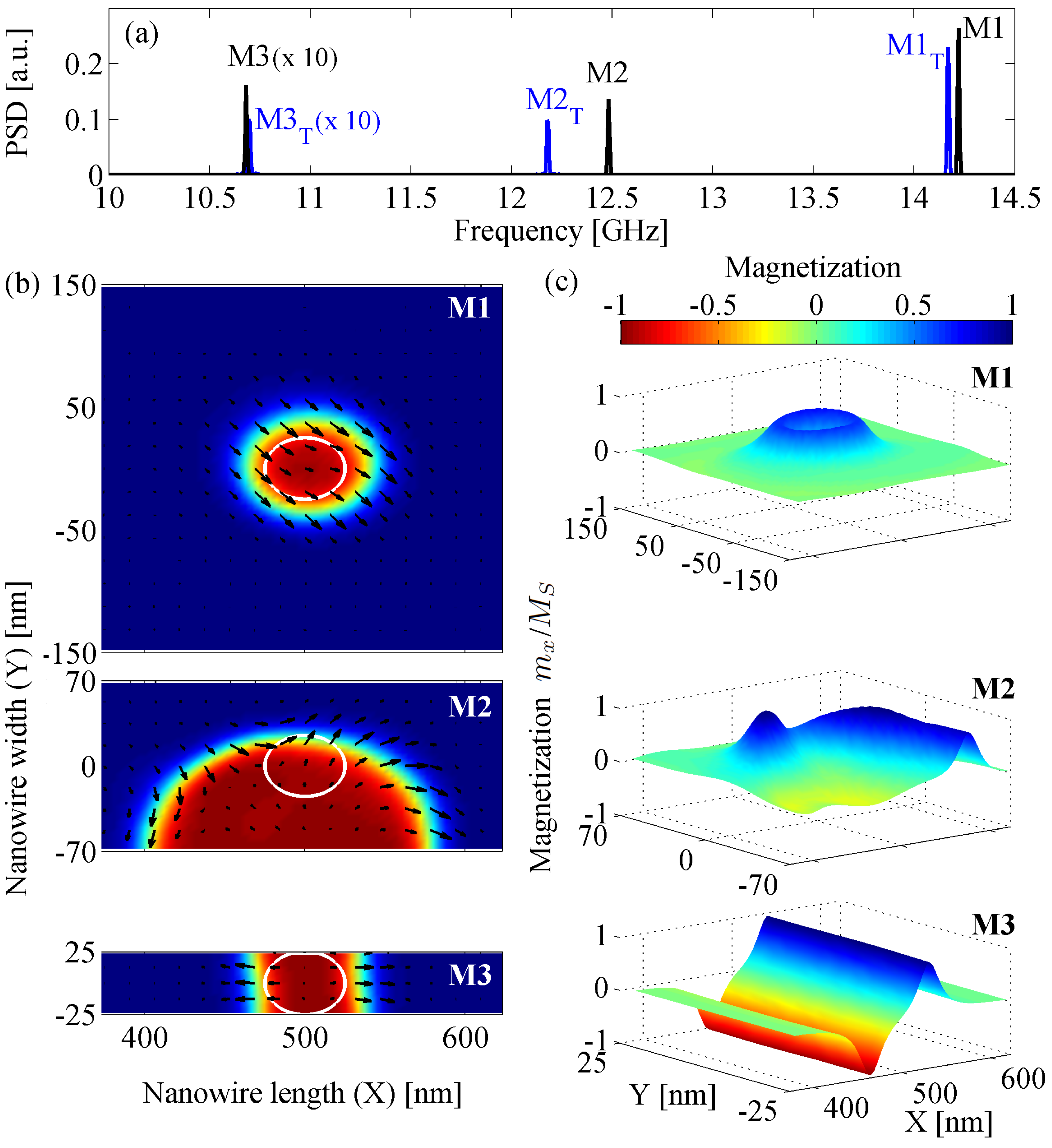}
\caption{ \label{fig3} (a) Spectra of the droplet modes, selecting the nanowire widths $300$~nm (M1), $140$~nm (M2), and $50$~nm (M3). The spectrum of M3 is magnified 10 times for clarity. The subscript T indicates the inclusion of thermal fluctuations. (b) and (c) show spatial profiles for the, respectively, $m_z$ and $m_x$ components of selected nanowire widths. Arrow plots of the in-plane components are also shown in (b) for clarity. The NC position is highlighted in white. The Supplementary Videos S1 are animated versions of (b). }
\end{figure}

In the following, we investigate the characteristics of each mode by choosing three representative nanowire widths, namely $300$~nm (M1), $140$~nm (M2), and $50$~nm (M3). Their simulated spectra are shown in Fig.~\ref{fig3}(a), exhibiting a clean resonance as expected for droplets under a perpendicular external field. These modes exist even when temperature, parameterized as a random thermal field~\cite{Brown1963} equivalent to $300$~K, is included in the simulations. Such spectra are indicated by the subscript T in Fig.~\ref{fig3}(a). The origin of the observed frequency shift, predominantly for M2, is beyond the scope of this Letter. Noteworthy, in the absence of thermal fluctuations, a resonant type of M3 is obtained at a nanowire width of $120$~nm, identified as a numerical artifact~\cite{SuppMat2013}.

A distinction between the modes can be made from their spatial profiles, shown as contour plots of the $m_z$ component [Fig.~\ref{fig3}(b)] and surface plots of the $m_x$ component [Fig.~\ref{fig3}(c)]. These plots confirm that M1 is the 2D droplet discussed in Ref.~\onlinecite{Hoefer2010}, including the Oersted-induced asymmetric location with respect to the NC. However, below a critical width, M2 arises as an edge mode where the droplet's footprint increases and `sticks' to the nanowire's side.  The edge M2 can be understood as a half 2D droplet~\cite{SuppMat2013}, in analogy to previously studied conservative three-dimensional surface droplets~\cite{Bespyatykh1994}. Furthermore, the Skyrmion number~\cite{Moutafis2007,Moutafis2009} of M1 and M2 is $\mathcal{N}=0$ as predicted by theory, motivating a similar analytical treatment for M2, as we describe below.

For the following theoretical discussion, we neglect the symmetry breaking Oersted and long-range demagnetizing fields in order to derive an analytical description.  The boundary conditions accompanying Eq.~\eqref{eq:llgs} are $\partial \hat{m}/\partial n = 0$ where $n$ is an outward pointing normal.  Since the 2D droplet is azimuthally symmetric, i.e. $\hat{m}_{\mathrm{droplet}} = \hat{m}_{\mathrm{droplet}}(\rho)$, $\rho = \sqrt{x^2 + y^2}$, we immediately observe that $\partial \hat{m}_{\mathrm{droplet}}/\partial y|_{y=0} = 0$ along the droplet centerline.  This implies that a droplet situated with its centerline at the boundary of the half-plane $y > 0$ is a solution of Eq.~\eqref{eq:llgs} and can describe M2 for sufficiently wide nanowires due to exponential localization in $x$ and for $y > 0$. Consequently, an asymptotic analysis~\cite{Hoefer2010} can be used to derive the sustaining current, $\sigma(I)$, for which zero total energy loss (gain - loss balance) is enforced, as required by all dissipative solitons
\begin{equation}
  \label{eq:edgecurr}
  \begin{split}
    \frac{\sigma(I)}{\alpha} = &- (H_k/M_s -1)[ \omega' + H_a/(H_k-M_s)] \\
    &\times \frac{\int_{y' > 0} \sin^2 \Theta(\vec{x}') d\vec{x}'}{\displaystyle \int\limits_{x'^2+(y'-\frac{\scriptstyle w'}{\scriptstyle 2})^2 < \frac{\scriptstyle d'^2}{\scriptstyle 4}} \frac{\sin^2 \Theta(\vec{x}')}{1+\nu \cos \Theta(\vec{x}')} d\vec{x}'},
  \end{split}
\end{equation}
where it is convenient to use primed, nondimensional coordinates with lengths scaled by $\lambda_{ex}/\sqrt{H_k/M_s-1}$ and frequencies scaled by $\gamma \mu_0 M_s^2/(H_k-M_s)$. The droplet profile $\Theta = \Theta(\rho')$ and shifted frequency $0 < \omega' < 1$ correspond to the ground state of the conservative 2D droplet equation $\omega' \sin \Theta = -\Theta'' - \Theta'/\rho' + \frac{1}{2} \sin 2 \Theta$~\cite{Kosevich1990}.  Equation~\eqref{eq:edgecurr} gives the relationship between the sustaining current and frequency.  The key difference between M2 and a 2D droplet is the denominator in Eq.~\eqref{eq:edgecurr} where the integral of the droplet is taken over the NC that is \emph{offset} from the droplet centerline.  Increasing the nanowire width $w'$ while keeping the NC centered corresponds to a shift of the NC further away from the footprint of M2.  Therefore, to maintain a fixed sustaining current, a wider nanowire requires a larger droplet footprint, precisely what is observed in Fig.~\ref{fig3}(b).  Furthermore, it is known that wider droplets exhibit lower frequencies~\cite{Hoefer2010}, explaining the frequency jump at the transition from M1 to M2 [Fig.~\ref{fig2}(b)].  Since the threshold current is not an intrinsic property of the resulting mode, as evidenced by the $3.5\%$ current variation at the M1-M2 transition [Fig.~\ref{fig2}(a)], we are justified in evaluating Eq.~\eqref{eq:edgecurr} and its 2D droplet counterpart~\cite{Hoefer2010} for the fixed current $I = -5.7$~mA which provides \emph{quantitative} agreement in frequency with the simulation results [solid red lines in Fig.~\ref{fig2}(a)].

Similar to 2D droplets, the existence of M2 is limited by a maximum sustaining current (recall that $I < 0$). Assuming negligible confinement effects, we theoretically determine the maximum sustaining current for M2 to fall below $-5.7$~mA for nanowires wider than $166$~nm, almost precisely the width at which the M1 - M2 transition occurs [Fig.~\ref{fig2}]. However, theory also predicts a higher, width-independent maximum sustaining current for M1 ($-3.3$~mA) so that mode selection is not completely explained by this argument. A detailed stability analysis is required to understand the preferred selection of M2.

We now turn our attention to M3. The corresponding contour plot of Fig.~\ref{fig3}(c) shows that magnetic boundaries are established only \emph{along} the nanowire's length, so that $\mathcal{N}=0$ (the unit sphere is not covered) and we identify M3 as a quasi-1D droplet. Additionally, the corresponding surface plot of Fig.~\ref{fig3}(c) shows that, as the soliton is traversed spatially, the magnetization vector undergoes a $360^\circ$ rotation. To understand the implications of the quasi-1D droplet features -- sub-Zeeman frequency and the magnetization vector ``twist'' -- we rely on known 1D droplet soliton solutions for uniaxial and biaxial materials in the absence of damping, STT, and symmetry breaking terms~\cite{Kosevich1990} i.e., when the underlying model is integrable. Only solitons with topological structure exhibit sub-Zeeman frequencies~\cite{Kosevich1990,Braun2012}, in contrast to M1 and M2 discussed above. Such a 1D topology is determined by the vector chirality~\cite{Braun2012}, defined as $\mathcal{\vec{C}}=\pi^{-1}\int dx (\hat{m}\times\partial_x\hat{m})$, from which we obtain an oscillatory behavior in time with magnitude $|\vec{C}|=2$. This indicates that M3 is a soliton-soliton pair with \emph{dynamic} magnetic boundaries, periodically morphing between a N\'{e}el- and Bloch-like configuration.

\begin{figure}[t]
\centering\includegraphics[width=3.3in]{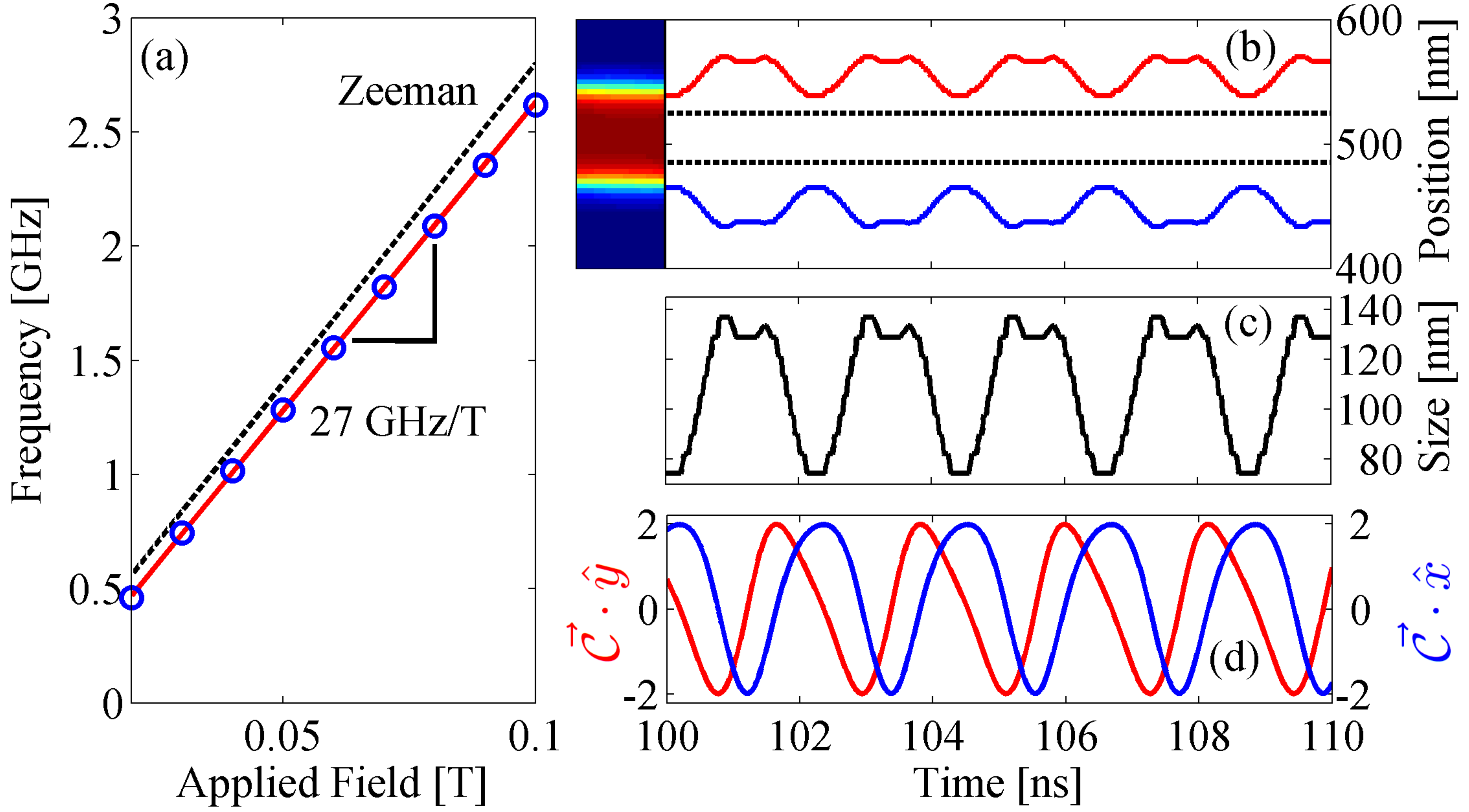}
\caption{ \label{fig4} (a) Quasi-1D droplet frequency as a function of field. The linear fit gives a $27$~GHz/T tunability. The Zeeman frequency is shown for comparison with black dashed line. (b) Position of the magnetic boundaries in time for $\mu_0H_a=0.02$~T, exhibiting oscillatory motion. The black dashed lines represent the NC edges. This motion leads to the quasi-1D droplet breathing (c) while the precession leads to the oscillatory change of the in-plane vector chirality (d). }
\end{figure}
To gain further insight, we perform field-dependent simulations on M3 nucleated in the nanowire of thickness $50$~nm and current $-3.5$~mA. As shown in Fig.~\ref{fig4}(a), M3 always exhibits a sub-Zeeman frequency. The field tunability can be obtained from a linear fit as $27$~GHz/T. Below $\mu_0H_a=0.02$~T, the quasi-1D droplet is nucleated but its oscillation frequency eventually relaxes to zero. An additional feature is observed as a function of field, namely, M3's spatial extent varies periodically in time at its precessional frequency. As an example, Fig.~\ref{fig4}(b) shows the averaged-in-$y$ position of the domain boundaries in time when $\mu_0H_a=0.02$~T, determined when $\langle m_z(t)\rangle_w=0$, where $\langle \cdot \rangle_w$ denotes averaging across the nanowire. The boundaries oscillate in ``anti-phase'', leading to a temporal variation, or breathing, of the quasi-1D droplet size [Fig.~\ref{fig4}(c) and Supplementary Video S2]. A breathing solution is provided by the integrable Landau-Lifshitz equation in the case of 1D biaxial ferromagnets in the zero field regime~\cite{Kosevich1990,Braun2012}. Such a solution can be described in terms of two bounded solitons in periodic motion. This picture agrees with the vector chirality discussed above whose $\hat{x}$ and $\hat{y}$ components are shown for this case in Fig.~\ref{fig4}(d).

The abovementioned similitudes between M3 and the integrable soliton solution are attributed to the increased \textbf{y} demagnetizing factor which promotes an effective biaxial anisotropy in the nanowire. For this reason, qualitative agreement is expected, such as the observed chirality and breathing. However, the inclusion of STT, damping, and field, break the integrability of the system. As a consequence, the breathing solution is only achieved by including a perpendicular field. Furthermore, the exact soliton solution breathes at twice the precessional frequency, in contrast with the simulated results. Consequently, an analytical treatment based on breather modes where damping is compensated by STT and a finite field is taken into account is required to compare with our results.

In summary, the 2D droplet undergoes mode transitions as a thin film is laterally reduced to the nanowire limit. There are primarily two novel modes observed: the edge droplet and the quasi-1D droplet. The former is an allowed solution of the existent dissipative droplet theory, which is evidenced by the excellent agreement between the micromagnetic simulations and the analytical estimates. On the other hand, the quasi-1D droplet is found to behave as a dynamical version of the breathing soliton-soliton pairs described in 1D biaxial ferromagnets. Interestingly, the quasi-1D droplet maintains a fixed vector chirality magnitude indicating that the soliton-soliton structure is stable, but its handedness periodically changes as the domain boundary precesses. Consequently, we believe that this novel droplet opens up new possibilities for low-dimensional droplet applications and magnetic soliton research.

Support from the Swedish Research Council (VR), the Swedish Foundation for Strategic Research (SSF), and the Knut and Alice Wallenberg Foundation is gratefully acknowledged. M. A. Hoefer gratefully acknowledges support from NSF CAREER DMS-1255422.

\section{References}
\bibliographystyle{aipnum4-1}

%

\end{document}